\documentclass[notitlepage,superscriptaddress,showpacs,nobalancelastpage,twocolumn,aps,prl]{revtex4-2}

\usepackage[english]{babel}
\RequirePackage[T1]{fontenc}
\RequirePackage{times} 

\usepackage{siunitx} 
\usepackage[italicdiff]{physics}
\usepackage{amsfonts}
\usepackage{lipsum}
\usepackage{multirow}
\usepackage{subfigure}
\usepackage{amsmath}
\usepackage{amssymb}
\usepackage{graphicx,epstopdf}
\usepackage{xspace}
\usepackage{array}
\usepackage[hidelinks]{hyperref}
\usepackage[dvipsnames]{xcolor}
\usepackage{my_symbols}
\usepackage{CJK}
\usepackage{bm}
\usepackage{verbatim}
\usepackage{tikz}
\usepackage{comment}
\usepackage{marginnote}
\usepackage[textwidth=1.5cm]{todonotes}
\usetikzlibrary{calc,3d}
\usetikzlibrary{arrows}
\usetikzlibrary{snakes}
\usepackage[normalem]{ulem}
\usepackage{soul}

\sethlcolor{green}

{\par}
\newcommand{\PT}{$\mathcal{PT}$}
\newcommand{\hmH}{\ensuremath{{\mathcal{H}}}}

\newcounter{mycomment}

\excludecomment{comment}  

\begin{document}

\begin{CJK*}{UTF8}{gbsn} 
\title{Antiferromagnetic Resonance Revisited: Dissipative Coupling without Dissipation}
\author{Yutian Wang}
\affiliation{Department of Physics and State Key Laboratory of Surface Physics, Fudan University, Shanghai 200433, China}
\author{Jiang Xiao (萧江)}
\email[Corresponding author:~]{xiaojiang@fudan.edu.cn}
\affiliation{Department of Physics and State Key Laboratory of Surface Physics, Fudan University, Shanghai 200433, China}
\affiliation{Institute for Nanoelectronics Devices and Quantum Computing, Fudan University, Shanghai 200433, China}
\affiliation{Shanghai Research Center for Quantum Sciences, Shanghai 201315, China}

\begin{abstract}
The antiferromagnet is a closed Hermitian system, we find that its excitations, even in the absence of dissipation, can be viewed as a non-Hermitian system with dissipative coupling. Consequently, the antiferromagnetic resonance spectrum does not show the typical level repulsion, but shows the level attraction --- a characteristic behavior often observed in non-Hermitian systems. Such behavior is because the antiferromagnetic ground state is \PT-symmetric. This new understanding on antiferromagnetic resonance also explains the mysterious enhancement of antiferromagnetic damping rate. Being effectively non-Hermitian, antiferromagnetic magnons can be used for quantum entanglement generation without introducing a third party like external pumping.


\end{abstract}

\maketitle
\end{CJK*}

\textit{Introduction.}
Ferromagnetic resonance (FMR) is the one of the most important phenomenon in the whole field of magnetism. It is widely used in experiments to characterize the physical properties of ferromagnetic materials. Its counterpart in antiferromagnets is the antiferromagnetic resonance (AFMR) \cite{kittel_theory_1948, keffer_theory_1952, nagamiya_antiferromagnetism_1955}, which happens in much higher frequencies of Terahertz (THz), in comparison with the Gigahertz (GHz) in FMR. As various types of THz source and detection methods become more common, the AFMR related effects attract more and more attention \cite{baltz_antiferromagnetic_2018}, such as the experimental demonstration of antiferromagnetic spin pumping due to AFMR \cite{cheng_spin_2014, li_spin_2020, vaidya_subterahertz_2020}.


The ferromagnet contains only one magnetic lattice, and the physics of FMR is qualitatively is identical to that for a driven oscillator. The most typical antiferromagnet is nothing but two opposite magnetic sublattices coupled via exchange interaction. And the AFMR in the collinear antiferromagnet has been well studied by Kittel \etal in 1950s \cite{kittel_theory_1948, keffer_theory_1952}, who illustrated that the eigenmodes in AFMR are two degenerate modes with left- and right-circular polarization. For such a simple and old effect, it seems that there is nothing new for understanding. However, we shall see that the AFMR spectrum cannot be regarded as two FMR coupled together, but have some peculiar feature that has been overlooked.

\begin{figure}[b]
  \includegraphics[width=\columnwidth]{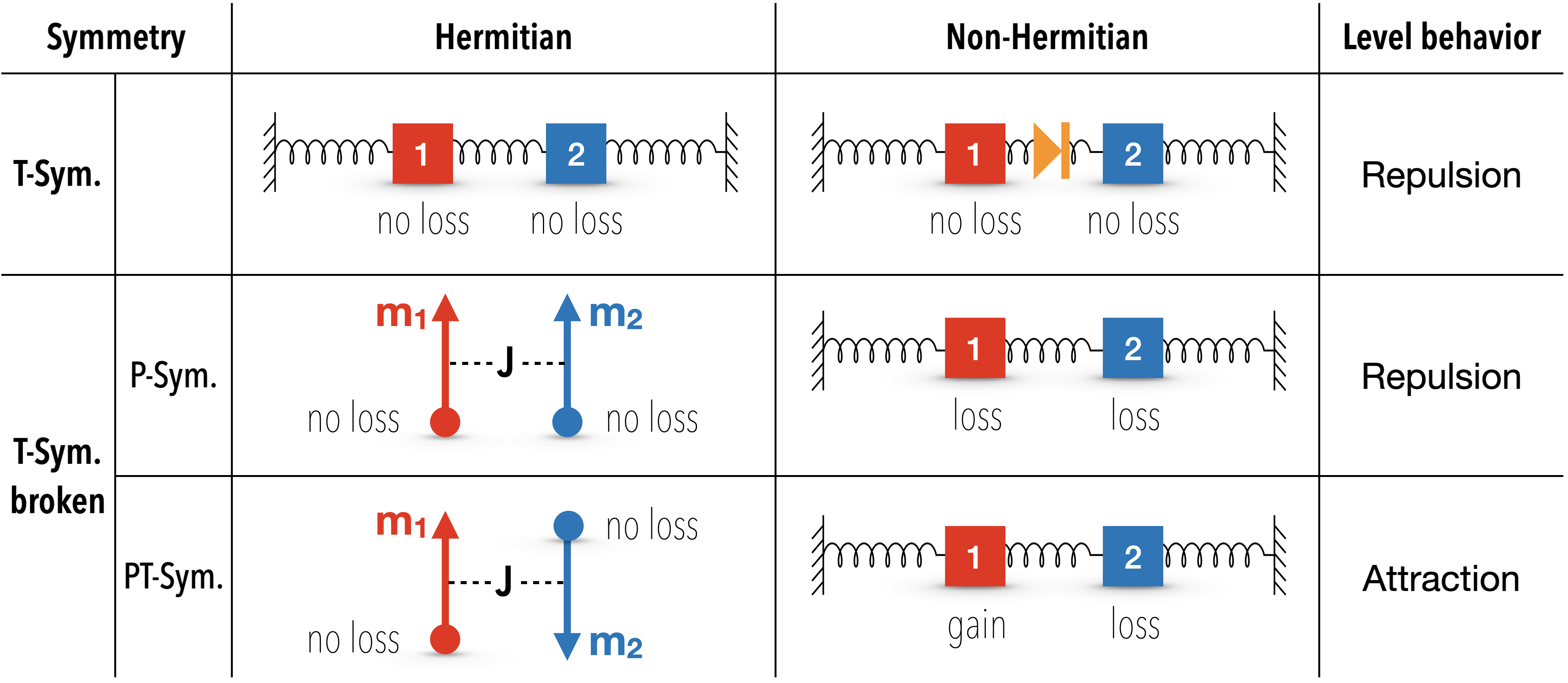}
  \caption{The representative Hermitian and non-Hermitian coupled systems. Row 1: Two lossless oscillators coupled via a conventional spring and a fictitious force diode are both \cT-symmetric. Row 2: The bipartite magnet with FM ground state and the coupled oscillators with loss both have \cT-symmetry broken, but retain the parity symmetry. Row 3: The bipartite magnet with AF ground state and the coupled oscillators with balanced gain and loss, being Hermitian and non-Hermitian respectively, are both \PT-symmetric and show the level attraction behavior.
}
  \label{fig:map}
\end{figure}

It is well known from the elementary quantum mechanics \cite{griffith_introduction_2005} that any coupling between two modes in a closed system should increase their frequency difference, known as the anti-crossing, level repulsion, or gap opening.
The antiferromagnet is generally regarded as a coupled system and its behavior should fall into such expectations. In this Letter, we shall see that this common wisdom does not hold for antiferromagnet: instead of level repulsion, the AFMR eigenmodes tend to attract each other.
The striking point is that such gap closing behavior is typically seen in non-Hermitian systems with the parity-time (\PT) symmetry (see \Figure{fig:map}) \cite{el-ganainy_non-hermitian_2018, ashida_non-hermitian_2020}, but the antiferromagnet in this discussion is Hermitian. Typical means of constructing such non-Hermitian systems include: i) introducing an equal gain and loss in the two subsystems so that the system becomes \PT-symmetric \cite{el-ganainy_non-hermitian_2018, ashida_non-hermitian_2020},
 and ii) coupling the two subsystems via the so-called dissipative coupling instead of the usual coherent coupling \cite{tserkovnyak_nonlocal_2005, yang_anti-mathcalpt_2017, harder_level_2018, grigoryan_synchronized_2018, yu_prediction_2019, yao_microscopic_2019, wang_dissipative_2020, tserkovnyak_exceptional_2020, troncoso_cross-sublattice_2021}.
One common feature shared by these two methods is the dissipation.

In this Letter, we point out that the dissipative coupling is not limited to non-Hermitian (or open) systems, but can also appear in closed Hermitian systems. And antiferromagnet is one such example, realizing 'dissipative' coupling without dissipation. The essential reason is that the antiferromagnetic ground state is \PT-symmetric (see \Figure{fig:map}).

\begin{figure*}[ht]
  \includegraphics[width=\textwidth]{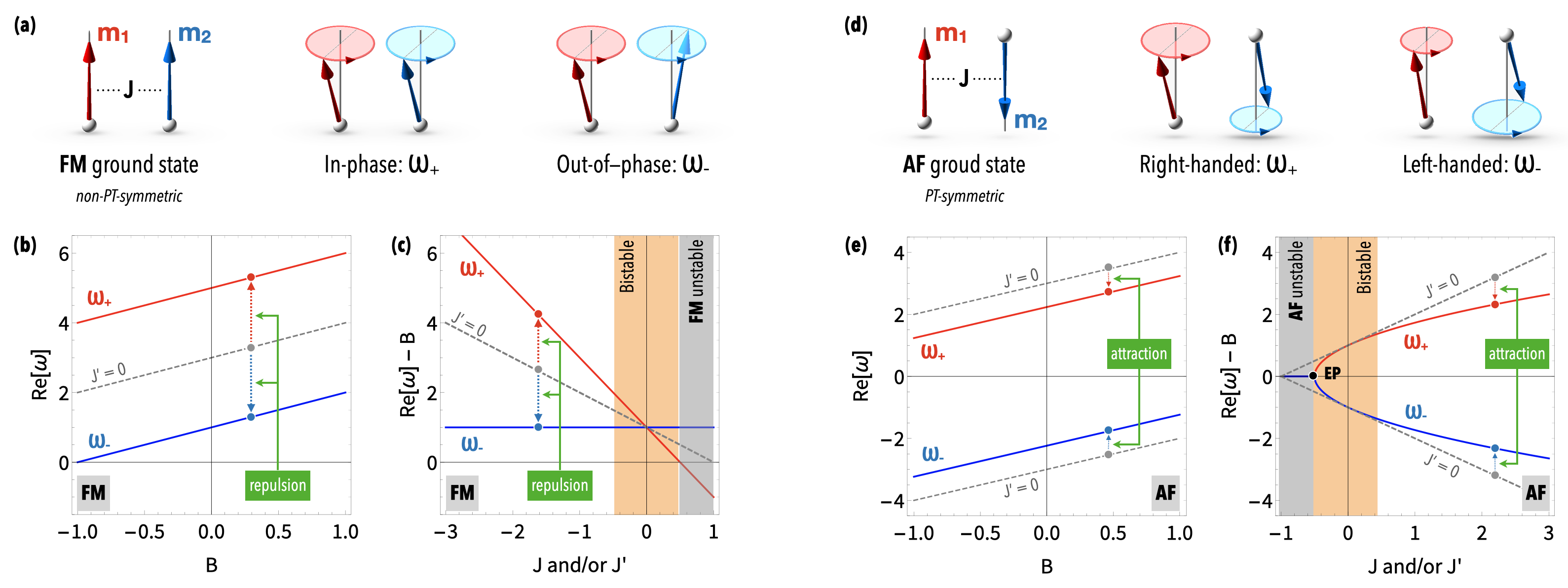}
  \caption{The resonance modes and their eigenfrequencies for FM and AF.
  (a) The FM ground state for bipartite magnet. The in-phase and out-of-phase eigenmodes upon FM ground state.
  (b) The field dependence of the eigenfrequencies in FM.
  (c) The coupling strength dependence of the eigenfrequencies (offset by $B$) in FM.
  (d) The AF ground state for bipartite magnet. The right-handed and left-handed eigenmodes upon AF ground state.
  (e) The field dependence of the eigenfrequencies in AF.
  (f) The coupling strength dependence of the eigenfrequencies (offset by $B$) in AF.
  All quantities have dimension of frequency and are in the unit of $K$. The solid and dashed curves in (b,c,e,f) are for including (and setting $J' = J$) and excluding ($J' = 0$) the transverse coupling.
  }
  \label{fig:wJJ}
\end{figure*}

\textit{Collinear Bipartite Magnet.} To illustrate the physics more clearly, we consider the simplest collinear bipartite magnet with two collinear magnetic sublattices as shown in \Figure{fig:wJJ}(a,d). The magnetic moments of the two sublattices, denoted by unit vectors $\mb_1$ and $\mb_2$, are coupled via the Heisenberg exchange. The moments are decomposed into the longitudinal $z$ component and the transverse $x$-$y$ components: $\mb_j = (m_j^x, m_j^y, m_j^z) = (\mb_j^\perp, m_j^z)$ with $j = 1, 2$. The Hamiltonian for the bipartite magnet is
\begin{equation}
  \label{eqn:H}
  \hH = -\sum_{j=1}^2 \qty(\frac{K}{2} {m_j^z}^2 + Bm_j^z)
  + J m_1^z m_2^z
  + J' \mb_1^\perp \cdot \mb_2^\perp,
\end{equation}
where $K$ is the strength of the uniaxial anisotropy along $\hbz$ and $B$ is the external magnetic field applied along $\hbz$. Here the Heisenberg exchange coupling between the two magnetic sublattices are distinguished as the longitudinal coupling $J$ along $\hbz$ and the transverse coupling $J'$ in the $x$-$y$ plane. Such distinction is to present the very different roles played by $J$ and $J'$. In most cases, we shall assume isotropic coupling with $J' = J$.
The coupled bipartite magnet favors antiferromagnetic ground state ($m_1^z = -m_2^z = 1$) when $J, J' > 0$. Otherwise, ferromagnetic ground state ($m_1^z = m_2^z = 1$) is favored. For $-0.5K<  J = J' < 0.5K$, both FM and AF ground states are stable due to the uniaxial anisotropy.
We set $\hbar = 1$ throughout the paper, so parameters $K, B, J, J'$ are all in the dimension of frequency.


{\it Ferromagnetic and Antiferromagnetic Resonance.} The dynamics of $\mb_{1,2}$ are governed by the coupled Landau-Lifshitz-Gilbert (LLG) equations:
\begin{align*}
 \dot{\mb}_1 &= -\mb_1\times\qty[\qty(B+K m_1^z-J m_2^z)\hbz-J'\mb_2^\perp - \alpha\dot{\mb}_1], \\
 \dot{\mb}_2 &= -\mb_2\times\qty[\qty(B+K m_2^z-J m_1^z)\hbz-J'\mb_1^\perp - \alpha\dot{\mb}_2],
\end{align*}
where $\alpha$ is the phenomenological Gilbert damping constant. Assuming $|\mb_j^\perp|\ll 1$ and $m_j^z\simeq \pm 1$, the resonance frequencies for the FM and AF ground states are solved as
\begin{subequations}
  \label{eqn:wpm}
\begin{align}
  \label{eqn:wpmFM}
  \omega_\ssf{FM}^\pm
  &= (1+i\alpha)(B + K -J \pm J'), \\
  \label{eqn:wpmAF}
  \omega_\ssf{AF}^\pm
  &= \qty[\pm\sqrt{(K+J)^2- J'^2} + i\alpha (K+J)](1\pm \td{B}),
\end{align}
\end{subequations}
with $\td{B} \equiv B/\sqrt{(K+J)^2- J'^2}$.
When setting $J' = J$, the expressions in \Eq{eqn:wpm} are well known for many decades. For the FM case, the two eigenmodes correspond to the in-phase and out-of-phase precession (see \Figure{fig:wJJ}(a)) of $\mb_1$ and $\mb_2$. For the AF case, the two eigenmodes are the left-handed and right-handed circular modes (see \Figure{fig:wJJ}(d)) \cite{keffer_theory_1952, cheng_spin_2014}.




The separation of $J$ and $J'$ in \Eq{eqn:wpm} indicates that the transverse coupling $J'$ is qualitatively different from the longitudinal coupling $J$.
Furthermore, the same transverse coupling $J'$ actually has opposite effects on the spectrum of FM and AF: in FM, the $J'$-coupling gives rise to the frequency gap between the two eigen modes
\footnote{The level repulsion and attraction is in terms of eigen frequency, not the eigen-energies (the absolute value of frequencies).
},
the same as in most coupled systems; while in AF, the $J'$-coupling reduces the frequency gap.
In \Figure{fig:wJJ}(b,c)/(e,f), we compare the scenarios with and without the $J'$-coupling, where we see the repulsive/attractive effect of the $J'$-coupling in FM/AF.
This level attraction or gap closing effect in AF resembles the similar behaviors observed in non-Hermitian systems \cite{yang_anti-mathcalpt_2017, harder_level_2018, grigoryan_synchronized_2018, yu_prediction_2019, yao_microscopic_2019, wang_dissipative_2020, tserkovnyak_exceptional_2020, troncoso_cross-sublattice_2021}.
Such non-Hermitian systems are required to have either the \PT-symmetry with balanced gain and loss or the dissipative coupling via an open or dissipative channel. However, the bipartite AF studied here is a Hermitian system that is closed and dissipationless. Therefore, the transverse coupling in AF, being non-dissipative, actually realizes the effect of dissipative coupling.
The two eigenfrequencies in AF coalesce at $J' = J = - K/2$ (see \Figure{fig:wJJ}(f)), known as the exceptional point. Beyond this point, the AF ground state becomes unstable, entering the \PT-symmetry broken region.



The bipartite magnetic system here is Hermitian and closed, but its excitations upon the AF ground state is actually governed by an effective non-Hermitian Hamiltonian \cite{ashida_non-hermitian_2020}.
In terms of complex fields $u_j = (m_j^y + i m_j^x)/\sqrt{2}$, the coupled LLG equation can be written as a Schr\"{o}dinger-like equation:
\begin{equation}
  i\pdv{t}\mqty(u_1 \\ u_2) = \hmH_{\rm eff} \mqty(u_1 \\ u_2)
  \label{eqn:Heff}
\end{equation}
with
\begin{equation*}
  \hmH_{\rm eff} = \begin{cases}
    B + (K-J)~~~~ +~~J'\hsigma_x &\qfor \mbox{FM} \\
    B + (K+J)\hsigma_z + iJ'\hsigma_y &\qfor \mbox{AF}
  \end{cases}
\end{equation*}
where $\hsigma_{x,y,z}$ are the Pauli matrices.
Clearly, $\hmH_{\rm eff}$ for FM is Hermitian, and the transverse coupling $J'\hsigma_x$ is of the coherent coupling type, which leads to the conventional repelling of the two degenerate eigenmodes as seen in \Figure{fig:wJJ}(b,c).
On the other hand, $\hmH_{\rm eff}$ for AF is non-Hermitian. But it is $\hsigma_z$-pseudo Hermitian, \ie $\hsigma_z\hmH_{\rm eff}\hsigma_z^{-1} = \hmH_{\rm eff}^\dagger$, thus it has real eigenvalues (for $J=J'>-K/2$) as given by the real part of \Eq{eqn:wpmAF} \cite{ashida_non-hermitian_2020}.
Here, the transverse coupling term $iJ'\hsigma_y$ has the form of dissipative coupling, and causes the attraction of the two eigen frequencies in AF seen in \Figure{fig:wJJ}(e,f).

As its name suggests, the dissipative coupling is usually associated with some kind of dissipative character \cite{wang_dissipative_2020}. For instance, the system itself is an open system with energy gain and/or loss \cite{harder_level_2018, yao_microscopic_2019}, or the coupling is mediated by a dissipative or leaking mode \cite{yu_prediction_2019, yang_anti-mathcalpt_2017}.
The AF system here is a closed system with no intrinsic dissipation (the Gilbert damping can be turned off), therefore the AF system realizes 'dissipative' coupling without dissipation.

\Figure{fig:wJJ}(f) also shows that the level attraction in AF actually happens for both antiferromagnetic coupling ($J >0$) and ferromagnetic coupling ($J<0$), as long as the system itself is in the antiferromagnetic ground state ($J> -0.5K$). This suggests that the dissipative coupling in AF has nothing to do with the antiferromagnetic coupling, but only relies on the AF ground state itself.
Since the bipartite magnet at AF ground state has the \PT-symmetry (see \Figure{fig:map}), we postulate that the excitations in a Hermitian system upon a \PT-symmetric ground state (or vacuum) always have an effective Hamiltonian that is pseudo-Hermitian, hence possess the level attraction feature in its spectrum.



{\it Quantum Description.} With Holstein-Primakoff transformation \cite{stancil_spin_2009}, we may rewrite $u_j$ as operators. To ensure the Bosonic commutation relation $[\ha,\ha^\dagger] = 1$, we make the substitutions: $u_j \ra \ha_j$ for the FM ground state, while $u_1 \ra \ha_1$ and $u_2 \ra \ha_2^\dagger$ for the AF ground state. This difference is because
$[m_1^y,m_1^x] = [m_2^y,-m_2^x] \simeq i$ in AF, but $[m_1^y,m_1^x] = [m_2^y,m_2^x] \simeq i$ in FM.
Using the creation and annihilation operators for the two sublattices, we rewrite the original Hamiltonian \Eq{eqn:H} (for $B=0$) as \cite{anderson_approximate_1952, rodriguez_linear_1959, anderson_concepts_1963}
\begin{equation}
  \hH = (K\mp J)\qty(\ha_1^\dagger\ha_1 + \ha_2^\dagger\ha_2)
  + J' \begin{cases}
  \ha_1 \ha_2^\dagger + \ha_1^\dagger\ha_2 \\
  \ha_1\ha_2 + \ha_1^\dagger \ha_2^\dagger
  \end{cases},
\label{eqn:Haa}
\end{equation}
where the upper/lower sign and line is for FM/AF, respectively.
It is evident that the $J'$-coupling term in FM state is particle-number conserving, but in AF state the $J'$-coupling becomes a particle-number non-conserving one. This feature can be traced back to the angular momentum conservation: in FM, the magnon excitation on both sublattices carries the same angular momentum of $-\hbar$ (along $\hbz$), therefore the creation of one magnon in one sublattice must be accompanied by the annihilation of one magnon in the other sublattice. In AF, on the other hand, the magnon excitations in the two sublattices carry opposite angular momentum, therefore the creation (or annihilation) of two magnons on two sublattice conserves angular momentum.
It can be shown (see SM) that the coupling between two mutual-time-reversal subsystems must be particle-number non-conserving, which further results in the non-Hermiticity for the dynamical excitations upon such \PT-symmetric Hermitian systems.
The connection between the particle-non-conserving terms and the non-Hermitian physics is also pointed out by Wang \etal \cite{wang_non-hermitian_2019}. Here we found that AF is a natural system to realize such particle-non-conserving term in a Hermitian way.
Because of this term, the antiferromagnet can be used to generate entanglement without the need of external parametric pumping, as demonstrated by Yuan \etal  \cite{yuan_enhancement_2020} and Hartmann \etal \cite{hartmann_intersublattice_2021}.

{\it Analogy with the Mechanical Oscillators.} It is illustrative to make an analogy between the bipartite magnet and the coupled harmonic oscillators, whose position and momentum are ($x_j, p_j$). The $m_j^x$ and $m_j^y$ components play the role of momentum and position of $i$-th oscillator. Their commutation relation are similar: $[m_j^y, m_j^x] = im_j^z \simeq \pm i$ versus $[x_j, p_j] = i$.
To keep the commutation relation exactly the same, we can make two types of mapping between the bipartite magnet and mechanical oscillators as the following:
\renewcommand{\arraystretch}{1.1}
\begin{center}
\begin{tabular}{
  >{\centering\arraybackslash}m{1.8cm} |
  >{\centering\arraybackslash}m{1cm} |
  >{\centering\arraybackslash}m{1cm}
  >{\centering\arraybackslash}m{1cm} |
  >{\centering\arraybackslash}m{1cm}
  >{\centering\arraybackslash}m{1cm}
  }
\multicolumn{2}{c|}{Bipartite Magnet} & \multicolumn{2}{c|}{Sublattice 1} &  \multicolumn{2}{c}{Sublattice 2} \\
  \multicolumn{2}{c|}{Ground State}& $m_1^x$ & $m_1^y$ & $m_2^x$ & $m_2^y$ \\ \hline
  \multirow{2}{*}{Mapping 1} & FM & $+p_1$ & $+x_1$ & $+p_2$ & $+x_2$ \\
                             & AF & $+p_1$ & $+x_1$ & $-p_2$ & $+x_2$ \\ \hline
  \multirow{2}{*}{Mapping 2} & FM & $+p_1$ & $+x_1$ & $-x_2$ & $+p_2$ \\
                             & AF & $+p_1$ & $+x_1$ & $-x_2$ & $-p_2$
\end{tabular}
\end{center}
Assuming the excitation is weak $\abs{m_j^{x, y}} \ll \abs{m_j^z} \sim 1$,
the bipartite magnet Hamiltonian \Eq{eqn:H} (at $B = 0$ and $J' = J$) is mapped via Mapping 1 as two coupled harmonic oscillators with equal natural frequency $\omega_1 = \omega_2 = K$:
\begin{equation}
  \hH
  = {K\ov 2} \sum_j (x_j^2 + p_j^2)
  +{J\ov 2}\begin{cases}
  - \qty[(x_1 - x_2)^2 + (p_1-p_2)^2] \\
  + \qty[(x_1 + x_2)^2 + (p_1-p_2)^2]
  \end{cases},
  \label{eqn:H0xp}
\end{equation}
where the upper/lower line is for FM/AF, respectively.
This suggests that the bipartite magnet with FM ground state is equivalent to the conventional coupled oscillators with coherent coupling term as $x_1x_2+p_1p_2$. While bipartite magnet with AF ground state can be regarded as an oscillator coupled to another oscillator that is in its time-reversal with $(x_2, p_2)\ra (x_2,-p_2)$, \ie coupling two mutual-time-reversal subsystems with dissipative coupling term as $x_1x_2 - p_1p_2$.

\begin{figure}[t]
  \includegraphics[height=5.0cm]{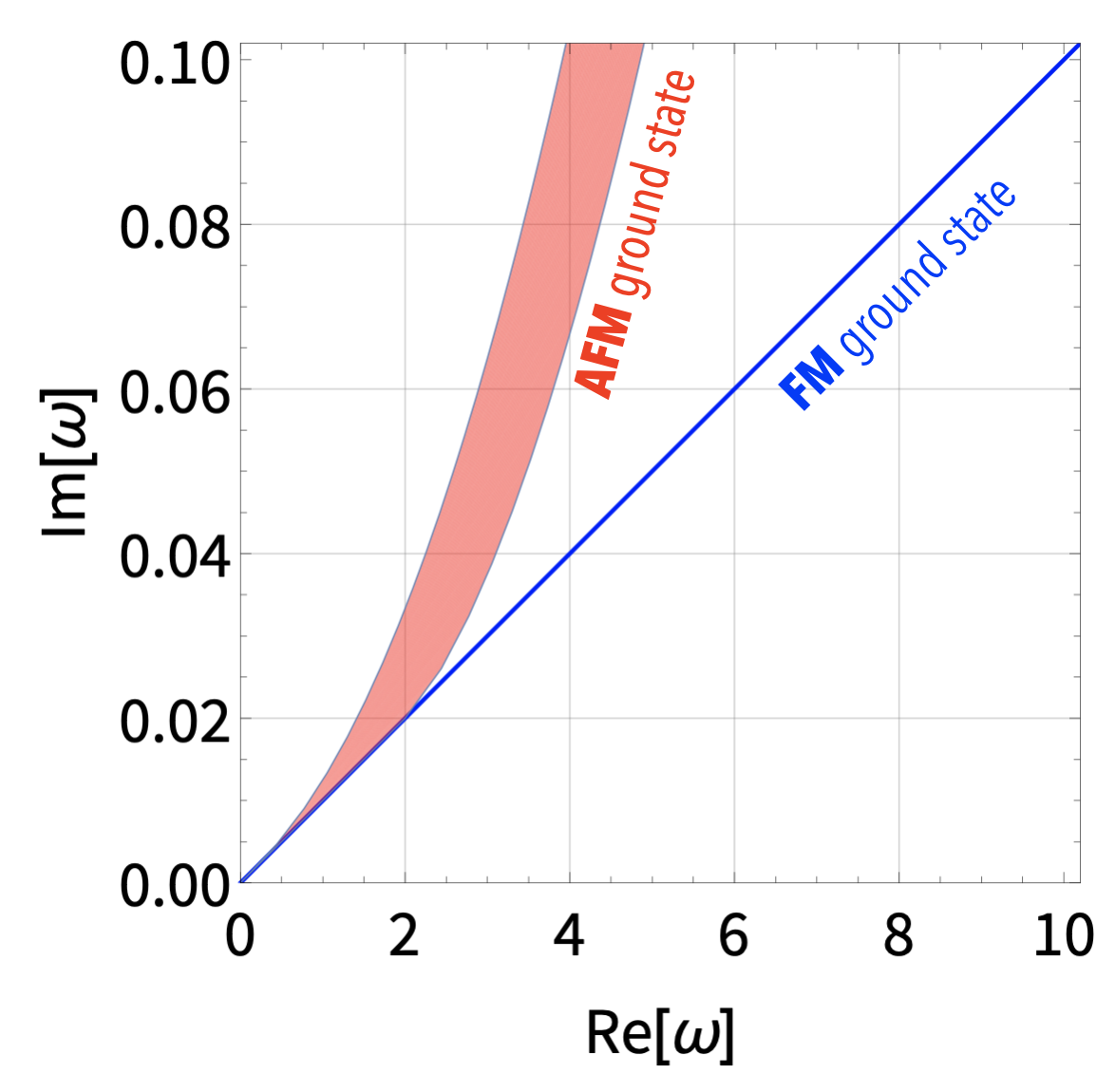}
  \caption{ The imaginary part of the eigenfrequency vs. the real part for FM (blue) and AF (red), as varying external field $B$ and/or exchange coupling $J, J'$. For FM, the damping rate is a constant $\alpha = 0.01$. For AF, it is a varying quantity larger than $\alpha$.
  }
  \label{fig:damping}
\end{figure}

{\it Enhanced Damping Rate in Antiferromagnet.}
Another unique feature in AFMR is the mysterious enhancement of its damping rate, defined as the ratio between the imaginary and real parts of the eigenfrequency. For coupled oscillators, the damping rates are limited by the largest damping parameter of individual oscillators. This is also the case in FM, \ie the damping rate in FM is given by the Gilbert damping parameter as seen in \Eq{eqn:wpmFM}: $\Im{\omega}/\Re{\omega} = \alpha$, independent of $J, J'$ or $B$.
In contrast, AF does not following this rule, and its damping rate is not a constant, and can far exceed the Gilbert damping constant $\alpha$ of each sublattice (see \Figure{fig:damping}). This damping rate enhancement in AF was also pointed out by Kamra \etal \cite{kamra_gilbert_2018}, but its understanding is still lacking.
With the attractive behavior of the eigenfrequencies in AF illustrated above, we understand now that the enhanced damping rate is in fact due to the reduction of the real part of the eigenfrequency, and there is no increase in the imaginary part at all.
This point can also be seen clearly from \Eq{eqn:wpmAF}, where the imaginary part of the frequency in AF is independent of the transverse coupling $J'$, and the real part is reduced by $J'$, consequently their ratio, the damping rate, increases because of $J'$. If the transverse coupling vanishes ($J' = 0$), the damping rate restores to the constant value $\alpha$. In other words, the transverse coupling $J'$ makes AF precessing slower than it should be, but leaving the dissipation unaffected.

{\it Discussion \& Conclusions.} We have only discussed a specific type of magnetic interaction --- the Heisenberg exchange coupling. The same physics can be applied to other type of magnetic interactions, such as the Dzayloshiinski-Moriya interaction (DMI) \cite{dzyaloshinsky_thermodynamic_1958, moriya_anisotropic_1960}.
As mentioned above, the level attraction behavior in AF is caused by its antiferromagnetic ground state, and does not correlate to the nature of the coupling. Therefore, the level attraction persists for the case of DMI as long as the ground state is AF.
The requirement for \PT-symmetric ground state can be relaxed to \PT-symmetric up to a shift or scaling,
which means that the application of external magnetic field or the uncompensated anti-ferrimagnet shall have similar features.

In conclusion, we found that the antiferromagnet, being a Hermitian physical system, can be regarded as an effective non-Hermitian system because of its antiferromagnetic ground state. Consequently, its spectrum or the AFMR has the level attraction behavior, which was previously believed to happen solely in non-Hermitian systems. We further postulate that the excitations upon a \PT-symmetric ground state shall possess the non-Hermitian physics, even if the system itself is Hermitian and closed. This new understanding of the antiferromagnetic spectrum also explains the mysterious enhancement of the damping rate in antiferromagnetic resonance. Furthermore, the intrinsic particle-number non-conserving Hamiltonian for AF can be extremely useful in quantum information, especially in entanglement generation.

{\it Acknowledgements.} This work was supported by the Science and Technology Commission of Shanghai Municipality (Grant No. 20JC1415900) and Shanghai Municipal Science and Technology Major Project (Grant No. 2019SHZDZX01).

\bibliographystyle{apsrev4-1}
\bibliography{ref}

\begin{comment}
\onecolumngrid
\appendix

\end{comment}

\end{document}